\documentclass[12pt]{article}

\usepackage{amsmath,multirow}
\usepackage{graphicx}
\usepackage{caption}
\usepackage{bm}
\usepackage{hhline}
\usepackage{booktabs}
\usepackage{enumerate}
\usepackage[authoryear]{natbib}

\newcommand{\mb}{\mathbf}
\newcommand{\bs}{\boldsymbol}

\newcommand{\mr}{\mathrm}
\newcommand{\mrE}{\mathrm{E}}
\newcommand{\Cov}{\mathrm{Cov}}
\newcommand{\mrM}{\mathrm{M}}

\def\mbu{\bm{u}}

\def\Var{\mr{Var}}

\def\mbf{\mb{f}}

\def\bb{\bm{\beta}}
\def\ba{\bm{\alpha}}

\def \mbQinv{\mb{Q}^{-1}}
\def \mbQ{\mb{Q}}

\def\mrB{\mr{B}}
\def\mrr{\mr{r}}
\def \mrP{\mr{P}}

\def \mbV{\bm{V}}
\def\bSig{\mb{\Sigma}}
\def\mbC{\bm{C}}
\def\rhoa{\rho_{\ba}}
\def \bV{\mb{V}}

\def\bI{\textbf{\text{I}}}
\def\mb{\bm}
\def\pr{^\prime}
\def\mrT{\mr{T}}
\def\mrN{\mr{N}}

\def\mbR{\mb{R}}

\def\mbU{\mb{U}}

\def\mbW{\mb{W}}

\def\mbK{\mb{K}}
\def\mbC{\mb{C}}
\def\mrJ{\mr{J}}

\def\mrI{\mr{I}}

\def\mrL{\mr{L}}
\def\mrM{\mr{M}}

\def\mrI{\mr{I}}
\def\mrm{\mr{m}}
\def\mrn{\mr{n}}

\def\b{2}
\def\c{1}

\DeclareMathAlphabet{\mathpzc}{OT1}{pzc}{m}{it}

\def\pr{^\prime}
\def\Cov{\mr{Cov}}

\def\bI{\mr{\textbf{I}}}

\def\mbf{\mb{f}}

\def\bb{\bm{\beta}}
\def\ba{\bm{\alpha}}

\def \mbQinv{\mb{Q}^{-1}}
\def \mbQ{\mb{Q}}

\def\mrB{\mr{B}}
\def\mrr{\mr{r}}
\def \mrP{\mr{P}}

\def \mbV{\bm{V}}
\def\bSig{\mb{\Sigma}}
\def\mbC{\bm{C}}
\def\rhoa{\rho_{\ba}}
\def \bV{\mb{V}}

\def\bI{\textbf{\text{I}}}

\def\mbB{\text{\textbf{B}}}
\def\mrB{\mr{B}}
\def \bpsi {\bm{\psi}}
\def \bPsi {\bm{\Psi}}

\def\gzt{ g\{Z_{\b i}(t)  \}}

\def\BB{BBSP}

\def\UB{UBSP}

\def\BF{BFPCA}

\def\UF{UFPCA}

\usepackage{amsmath}
\makeatletter
\renewcommand*\env@matrix[1][c]{\hskip -\arraycolsep
  \let\@ifnextchar\new@ifnextchar
  \array{*\c@MaxMatrixCols #1}}
\makeatother

\begin{document}

\begin{center}
	{\Large {\bf Modeling Multivariate Mixed-Response Functional Data}}\\
    Beth A. Tidemann-Miller\footnote{Biogen Inc., Cambridge, Massachusetts},
	Brian J. Reich\footnote{
		Department of Statistics, North Carolina State University, 
		Raleigh, North Carolina},
	and Ana-Maria Staicu$^2$\\
	\today
\end{center}

\begin{abstract}
We propose a Bayesian modeling framework for jointly analyzing multiple functional responses of different types (e.g. binary and continuous data). Our approach is based on a multivariate latent Gaussian process and models the dependence among the functional responses through the dependence of the latent process. Our framework easily accommodates additional covariates.  We offer a way to estimate the multivariate latent covariance, allowing for implementation of multivariate functional principal components analysis (FPCA) to specify basis expansions and simplify computation. We demonstrate our method through both simulation studies and an application to real data from a periodontal study. 

\end{abstract}

\section{Introduction}\label{s:intro}

Until recently, the primary focus of methods employing functional principal components analysis (FPCA) has been on real-valued functional responses. Methods that can model non-Gaussian functional responses, such as repeatedly observed binary or count data, are only recently appearing for univariate functional responses (for example: \cite{Hal+:08}, \cite{van:09}, \cite{Ser+:13}). Additionally, methods that extend functional modeling from the univariate case (i.e. one response curve) to the multivariate case (i.e. a vector of multiple response curves) are currently undergoing development (for example: \cite{Zho+:08}, \cite{Ber+:11}, \cite{Jac+Pre:14}). These multivariate functional methods are limited in that all curves comprising the multivariate response vector must be real-valued. 

Here we propose a Bayesian multivariate functional model that utilizes a multivariate latent Gaussian process and can handle responses of different types, e.g. binary and continuous data. Our method easily incorporates covariates, a feature previously unavailable for modeling non-Gaussian functional responses. As an extension of the methods of  \cite{Hal+:08}, we propose a way to estimate the multivariate latent covariance, in particular, the cross-covariance of latent functions corresponding to different responses. 
By using a reliable estimate of the multivariate latent covariance, our proposed method can implement multivariate FPCA to specify basis expansions and simplify computation.

Several approaches to modeling non-Gaussian univariate functional responses have appeared in the literature. For binary or count data observed repeatedly, \cite{Hal+:08} proposed a non-parametric functional approach in which the observed responses are directly related to a latent Gaussian functional process through a link function. In order to implement FPCA, they used a Taylor series approximation to derive estimators of the latent process mean function and covariance operator and used bootstrapping methods for further inference. A similar approach by \cite{Ser+:13} used logistic functional regression to model multilevel cross-dependent binary-valued functional data. In the case of non-rare events, their approach is an extension of the linear approximation methods of \cite{Hal+:08} to multilevel data. For rare events, they introduced an approach centered around an exponential approximation. 

In contrast to the aforementioned frequentist methods, \cite{van:09} offered a Bayesian approach to FPCA for repeatedly observed binary or count data. They extended the variational algorithm for Gaussian responses given in \cite{van:08}, and focused on canonical links for one-parameter exponential families.
The methods of \cite{Hal+:08}, \cite{Ser+:13} and \cite{van:09} offer ways to model univariate functional responses, whereas the approach we propose in this paper jointly models multivariate functional responses of mixed type. 

To date, the literature concerning multivariate FPCA has been sparse. \cite{Ram+Sil:05} gave a brief example that uses FPCA for a bivariate functional response of hip and knee angle measurements for gait data. After assigning the two functional responses to a fine grid of points, they concatenated the two response functions and proceeded with PCA in the traditional multivariate framework. \cite{Ber+:11} proposed multivariate FPCA in which the principal components are smooth functions, a result of performing FPCA at each observed location in a domain on which curves have been smoothed. In contrast to the approach of \cite{Ram+Sil:05}, \cite{Jac+Pre:14} presented a method that allowed for non-orthonormal bases which made it possible for each curve in the multivariate response vector to have its own basis expansion. Their approach neatly addresses how to handle responses with differing magnitudes of variation within the curves.

To our knowledge, our method that models multivariate mixed-type responses is the first of its kind within the functional data analysis literature. In the spatial literature, \cite{Rei+Ban:10} developed a spatial latent factor model for multivariate mixed-response data with informative missingness. Our approach shares several similarities to that of \cite{Rei+Ban:10}, however our approach is able to examine complex correlation structures that their stationary spatial method is not equipped to handle. 

%%%%%%%%%%%%%%%%%%%
\section{Model}
\subsection{General Framework}\label{s:model}

We present the following methodology to jointly model $\mrP$ functional responses. Denote $Y_{pi}(t)$ as the observed functional response of type $p=1,\hdots,\mrP$ for subject $i=1,\hdots,\mrN$ at location $t \in {\cal T}$. The responses $Y_{pi}(t)$ are observed only at a finite set of $\mrL_{pi}$ locations $t_{pi1},t_{pi2},\hdots,t_{pi\mrL_{pi}}$, which may be different for subject and response type. To combine responses with different supports, e.g., binary and continuous, let $Y_{pi}(t) = h_p \{ W_{pi}(t) \}$ for link function $h_p(\cdot)$ and latent response $W_{pi}(t)$. Motivated by the periodontal application in Section \ref{s:dental_data}, we restrict our attention to Gaussian and binary responses. If response $p$ is Gaussian then we use the identity link $h_p(\eta)=\eta$; if response $p$ is binary, then we use the indicator link $h_p(\eta)=\mrI(\eta>0)$.

Dependence between responses is modeled via the latent Gaussian processes
\begin{align} \label{latent1}
		W_{pi}(t) = Z_{pi}(t) + \epsilon_{pi}(t)
\end{align}
 where ${\epsilon}_{pi}(t) \overset{iid}{\sim} \mrN(0,\tau_p^2)$ is random noise and $Z_{pi}(t)$ is a random process. For identification purposes, we fix $\tau_p=1$ for binary responses. Furthermore, let $Z_{pi}(t) = \mu_{pi}(t) + f_{pi}(t)$, the sum of a fixed mean function $\mu_{pi}(t)$ and a smooth subject-specific process $f_{pi}(t)$, assumed to be uncorrelated with ${\epsilon}_{pi}(t)$.

The mean can be modeled as $\mu_{pi}(t)= \sum_{j=1}^{\mrm_{p}}x_{pij}(t) \beta_{1pj} + s_p(t)$ so that it can incorporate $\mrm_{p}$ covariates $x_{pij}(t)$ with fixed coefficients $\beta_{1pj}$ and a population-level smooth function $s_p(t)$. It is possible for a subject-specific covariate to depend on the functional location $t$, for example the indicator of jaw in the periodontal data of Section \ref{s:dental_data}, and it is also possible for the same covariates to affect all responses. The smooth function $s_{p}(t)$ is assumed to be square integrable on ${\cal L}^2[0,1]$. We use a predetermined basis expansion to approximate $s_p(t)$. Let $\{ {\mrB}_{p j}(t): 1 \le j \le \mrn_{p} \}$ be a basis expansion in ${\cal L}^2[0,1]$  of dimension $\mrn_{p}$. We approximate the smooth part by $s_p(t)= \sum_{j=1}^{\mrn_{p}} { {\mrB}_{p j}(t)\beta_{2p j}}$ where the type of basis expansions are allowed to differ across response $p$. To simplify notation, we write $\mu_{pi}(t)= \mbu^\mrT_{pi}(t) {\bb_p}$ where $\mbu_{pi}(t)=[x_{pi1}(t),\hdots,x_{pi\mrm_p}(t),{\mrB}_{p1}(t),\hdots,{\mrB}_{p \mrn_p}(t)  ]^\mrT$ is a vector of length $\mrJ_p=\mrm_{p}+\mrn_{p}$ that combines the covariates and basis functions and has corresponding coefficient vector ${\bb_p}=[ {\beta}_{1p1},\hdots, {\beta}_{1p \mrm_{p}}, {\beta}_{2p 1},\hdots, {\beta}_{2p \mrn_{p}}]^\mrT$.

Let  $\mb{f}_i(t)=[f_{1i}(t),\hdots ,f_{\mrP i}(t) ]^\mrT$  be the vector of random subject-specific deviation functions and assume $\mb{f}_i(t)$ are i.i.d. mean-zero Gaussian processes where  $\Cov\{\mb{f}_i(t),\mb{f}_i(t\pr)\}= \mbK(t,t\pr)$ and $K_{pp\pr}(t,t\pr)=\Cov \{f_{pi}(t),f_{p\pr i}(t\pr)\}$ form the elements of $\mbK(t,t\pr)$. The covariance operator $K_{pp\pr}(t,t\pr)$ captures both auto-dependence ($p=p\pr$) and cross-dependence ($p\neq p\pr$) between two different latent responses. We assume that $f_{pi}(t)$ is a  smooth process in ${\cal L}^2[0,1]$ and present two ways of specifying basis expansions for  $f_{pi}(t)$: Section \ref{s:pre} details how to use predetermined bases and Section \ref{s:datadriven} gives a data-driven approach that uses multivariate FPCA. 

We can write the multivariate model succinctly in matrix form. Let ${\bb}=[\bb_1^\mrT, \hdots, \bb_\mrP^\mrT ]$  be the fixed effect vector of length $\mrJ= \sum_{p=1}^\mrP  \mrJ_p $ with corresponding $\mrP \times \mrJ$ matrix $\mbU_i(t)$ comprised of appropriate evaluations of $\mbu_{pi}(t)$. Let $\mb{\epsilon}_i(t) \overset{iid}{\sim} N(0, \mb{D})$ where $\mb{D}$ is diagonal with elements $\tau_1^2,\hdots,\tau_\mrP^2$.
 Then \eqref{latent1} becomes \begin{align}
	\label{mod1}
		\mbW_i(t)=  \mbU_i(t) \bb + \mb{f}_i(t) + \mb{\epsilon}_i(t) .
	\end{align}

%-------------------------------------------------------------------------
	\subsection{Predetermined bases}\label{s:pre}
%-------------------------------------------------------------------------

The first way in which we specify basis expansions for $f_{pi}(t)$ is by choosing predetermined bases such as B-spline, Fourier, or polynomial bases.
Let 
\begin{align} \label{f1}
	f_{pi}(t)= \sum_{k=1}^{\mrM_p} \psi_{pk}(t) \alpha_{pik}
\end{align}	
	 where $\{ \psi_{p k}(t): 1 \le k \le \mrM_p \}$ is a basis expansion in ${\cal L}^2[0,1]$ of dimension $\mrM_p$ and $\ba_{pi} = [\alpha_{pi1},\hdots,\alpha_{pi{\mrM_p}} ]^\mrT$ are random coefficients with $\mrE(\alpha_{pik})=0$ and $\Cov(\alpha_{pik},\alpha_{p\pr  i \ell })= \xi_{k \ell p p\pr}$. The multivariate covariance function induced by \eqref{f1} is 
\begin{align}
	K_{pp\pr}(t,t\pr)=\Cov\{f_{pi}(t), f_{p\pr i}(t\pr)\} = \sum_{k=1}^{\mrM_p} \sum_{\ell=1}^{\mrM_{p\pr}} \psi_{p k}(t) \psi_{p\pr \ell}(t\pr)   \xi_{k \ell p p\pr},
\end{align}	
which is a function of both the basis functions and covariance $\bSig=\{ \xi_{k \ell p p\pr} \}$. 
	 Using predetermined basis expansions is extremely flexible; % in that it can approximate any covariance function. 
in the Appendix, we discuss how the covariance model can approximate the covariance matrix of any arbitrary finite-dimensional distribution. The choice of $\mrM_p$ is important in that one needs to select a number of basis functions that is sufficient to approximate the covariance well but is not unnecessarily large. We suggest choosing $\mrM_p$ based on a grid search, using criteria such as deviance information criteria (DIC) for comparison.

	\subsection{Data-driven bases}\label{s:datadriven}
As an alternative to using predetermined bases, we introduce a novel approach in which we use estimated basis functions that are obtained through FPCA of the multivariate latent covariance.
We propose FPCA for multivariate mixed-responses, inspired by \cite{Hal+:08} who introduced FPCA for binary-valued functional responses. We too require that the probability of observing a binary event is sufficiently far from zero or one. For simplicity of presentation, we ignore the covariates and discuss how to account for them later in this section.

Recall from \eqref{latent1} that we model the $p^{\mr{th}}$ response as $Y_{pi}(t) = h_p \{ W_{pi}(t) \}$ through the latent Gaussian process $W_{pi}(t) = Z_{pi}(t) + \epsilon_{pi}(t)$ and link function $h_p(\eta)$. Linking the latent response directly to the observed response is equivalent to assuming there is a corresponding monotone link function $g_p(\cdot)$ such that $\mrE\{Y_{ip}(t) | Z_{pi}(t) \} = g_p\{Z_{pi}(t)\}$; we focus on $g_p$ here.
Following \cite{Hal+:08}, assume that $g_p(\cdot)$ has bounded fourth derivative and that the latent process  satisfies $Z_{pi}(t)= \mu_{p}(t) + \delta X_{pi}(t)$ for fixed mean $\mu_p(t)$, unknown small constant $\delta>0$, and mean-zero Gaussian random variable $X_{pi}(t)$ that is i.i.d. across subjects $i$ and has both finite variance and finite covariance between $X_{pi}(t)$ and $X_{p\pr i}(t\pr)$. Our goal is to approximate the latent covariance matrix of $Z_{pi}(t)$ whose covariance operator is $K_{pp\pr}(t,t\pr)=\Cov\{Z_{pi}(t), Z_{p\pr i}(t\pr)\}$. Without loss of generality, we restrict our attention to one continuous Gaussian response ($p=1$) and one binary response ($p=2$) with link functions $g_1(\eta)=\eta$ and $g_2(\eta)=\Phi(\eta)$ where $\Phi(\cdot)$ is the standard normal cdf function. For simplicity, we use $g$ to denote $g_2$ in the following exposition.

The covariance consists of variance components $K_{pp}$ and cross-covariance components $K_{pp\pr}$. The variance components $K_{11}$ and $K_{22}$ are estimated using the common FPCA for continuous responses \cite{Ram+Sil:02, Ram+Sil:05} 
as well as binary-valued responses \citep{Hal+:08}, respectively. In particular, when the responses are binary valued, the variance $K_{22}$ is estimated using 
 \begin{align} \label{Hall_univ}
\widetilde{K}_{22}(t,t\pr) = \{\hat{S}_{22}(t,t\pr)- \hat{\eta}_2(t)\hat{\eta}_2(t\pr) \} /[ g^{(1)}\{\hat{\mu}_2(t)\}  g^{(1)}\{\hat{\mu}_2(t\pr)\}   ],
\end{align}
where $g^{(1)}$ indicates the first derivative of $g$. The latent mean estimator is $\hat{\mu}_p(t)=g^{-1}\{ \hat{\eta}_p(t) \}$ where $\hat{\eta}_p(t)$ estimates $\mrE[g\{ Z_{pi}(t) \}] ={\eta}_p(t)$ and is found by smoothing the data $\big(t,Y_{p i}(t)\big)$ for $i=1,\hdots,\mrN$. $\hat{S}_{22}(t,t\pr)$ is the estimator for $S_{22}(t,t\pr)= \mrE\{Y_{2 i}(t)Y_{2 i}(t\pr)\}= \mrE\big[ g\{ Z_{2i}(t) \} g\{ Z_{2i}(t\pr) \} \big]$ and is obtained through bivariate smoothing of the data  $\big((t,t\pr),Y_{2 i}(t) Y_{2 i}(t\pr) \big)$ for $i=1,\hdots,\mrN$, removing the diagonals before smoothing.

For the cross covariance operator $K_{12}$ we remark that 
\begin{align} \label{cross}
{K}_{12}(t,t\pr) = \Cov \big\{Y_{1 i}(t),Y_{2 i}(t\pr) \big \}/g^{(1)}\{\mu_2(t\pr)\},
\end{align}
which is obtained by approximating 
$ \Cov \big \{ Y_{1 i}(t),Y_{2 i}(t\pr) \big \}= \Cov \big[Z_{1i}(t),g\{Z_{2 i}(t\pr)\} \big]$ 
using a Taylor expansion of $g\{Z_{2 i}(t\pr)\}$ around $\mu_{2}(t\pr)$. More details are given in the Appendix. 
This leads to the estimator of the cross component given by 
\begin{align} \label{cross2}
\widetilde{K}_{12}(t,t\pr) = \{\hat{S}_{12}(t,t\pr)- \hat{\eta}_1(t)\hat{\eta}_2(t\pr) \} / g^{(1)}\{\hat{\mu}_2(t\pr)\}.
\end{align}

\sloppy
Combining the individually smoothed estimators $\widetilde{K}_{11}(t,t\pr)$, $\widetilde{K}_{22}(t,t\pr)$ and $\widetilde{K}_{12}(t,t\pr)= \widetilde{K}_{21}(t\pr,t)$ forms the smooth $2\times 2$ estimator $\widetilde{\mbK}(t,t\pr)$ of the bivariate latent covariance operator. Note that for smoothing purposes in this paper, we implement a global smoother as opposed to the local least squares smoothing of \cite{Hal+:08}, though either is appropriate. In the presence of subject-specific covariates, one can find covariate estimates using least squares or logistic regression, depending on the type of response, and then use the residuals to estimate the latent covariance.

The final step for creating basis functions is to implement bivariate FPCA in which we obtain the eigenfunctions ${\bs{\theta}}(t)=[{\theta}_1(t), \hdots, {\theta}_{\mrP}(t)]^\mrT$ and the eigenvalues ${\lambda}$ of the matrix $\widetilde{\mbK}(t,t\pr)$.  Note that the matrix $\widetilde{\mbK}(t,t\pr)$ is not guaranteed to be positive definite, but we can ensure the truncated spectral decomposition $\widetilde{\mb{K}}(t,t\pr)= \sum^{\mrM}_{k=1} {\lambda}_{k}{\bs{\theta}}_{k}(t) \{{\bs{\theta}}_{k}(t\pr)\}^\mrT$ is positive definite by restricting the inclusion of only positive eigenvalues and their eigenfunction counterparts.
The truncation value $\mrM$  is chosen based on the proportion of variation explained by the eigenvalues as suggested in \cite{Di+:09}.
In particular,  specify a cumulative explained variance threshold $\mr{P}_1$ and individual explained variance threshold $\mr{P}_2$. Define
$\mrM = \min \{k: p_{1k} \ge \mr{P}_1, p_{2k} < \mr{P}_2 \} $ where 
$p_{1k}=\sum_{i=1}^{k}{{{\lambda}_{i}}/	\sum_{j=1}^{n}{{ \lambda}_{j}}}$, $p_{2k}={\lambda}_k/	\sum_{j=1}^{n}{{ \lambda}_{j}}$ and the positive eigenvalues are the first $n\ge k$ eigenvalues. We specify the basis functions to be the eigenfunctions scaled by their associated eigenvalues, ${\psi}_{pk}(t) = \sqrt{{\lambda}_k} {{\theta}}_{pk}(t)$, and the subject-specific deviation function is approximated by 
$f_{pi}(t)= \sum_{k=1}^\mrM  {\psi}_{pk}(t) \alpha_{ik}$. 

Using this data-driven basis approach, the correlation across responses is largely captured by the basis functions from FPCA. Additionally, since each basis function combines information from all responses, the data-driven approach results in one set of basis functions, eliminating the need to have a set of basis functions for each response. These distinctions offer important advantages over the predetermined basis approach. First, having only one set of basis functions reduces the dimensionality of the random-effect covariance matrix $\bSig$, making it easier to fit. Second, it allows for further simplification since $\bSig$ is diagonal. This will offer computational advantages over the predetermined basis method where the burden of capturing the correlation across responses falls entirely on estimating a non-diagonal $\bSig$ which can potentially have very large dimension. 

One important consideration to make when implementing this data-driven basis function approach is to ensure that the variance of the latent process for the continuous component is on a scale similar to that of the latent process for the binary component. We suggest scaling the continuous process by $Y_{1 i}(t)/s$ where $s$ is the overall sample standard deviation of the continuous response without regard to $t$. Since $s$ is a scalar quantity, it is straightforward to scale prior to implementing the latent covariance, FPCA and MCMC estimation algorithms, rescaling only the final results back to the original scale. 
%----------------------------------------
\subsection{Prior Specification}\label{s:priors}
%----------------------------------------

To complete the Bayesian model, we specify priors for the hyperparameters.
The fixed effect parameters ${\bb}$ are assigned uninformative Gaussian priors. 
Let the subject random effect $\ba_i$ have a Gaussian prior with $\Cov(\ba_i)=\bSig$ and assign $\bSig$ an Inverse Wishart prior. For the error variances of the continuous processes, let $\tau^2_p$ have an uninformative gamma prior;  for identifiability $\tau^2_p$ is fixed at 1 for binary processes. In summary,
	\begin{align}
	\label{eqn:priors}
		&\bb | \sigma^2_\beta \sim \mrN_\mrJ(\mb{0},  \sigma^2_\beta\bI_\mrJ) \nonumber\\
		&\ba_i | \bSig \sim \mrN_\mrM(\mb{0}, \bSig) \nonumber \\
		&\bSig | q_1, q_2 \sim \text{InvWishart}_\mrM( \mbV=q_2\bI_{\mrM}, \nu=q_1)\\
		& \tau^2_p | l,h \sim \mr{InvGamma}(l,h) \nonumber
	\end{align}
for hyperparameters $\sigma^2_b$, $q_1$, $q_2$, $\ell$, and $h$, selected to result in weak priors.

\section{Computational Details}\label{s:compdetails}

To facilitate MCMC sampling, we treat the continuous latent processes $W_{pi}(t)$ for binary response as unknown parameters to be updated as part of the sampling as in \cite{Alb+Chi:93}. Using this auxiliary variable approach, all parameters have conditional conjugacy due to the prior specifications given in Section \ref{s:priors}, allowing us to implement Gibbs sampling. The Gibbs sampling algorithm uses full-conditional posteriors derived in the Appendix and which use notation that we now describe.

Denote the observation locations as $t_{pi\ell}$, $\ell = 1, \hdots, \mrL_{pi}$, for each subject $i$ and response $p$, giving a total of $\mrL_i =  \sum_{p=1}^\mrP \mrL_{pi}$ locations. Let $n=\sum_{i=1}^\mrN{\mrL_i}$ be the total number of locations observed across all subjects. 
Let $\mbW_{pi}$ be the vector of length $\mrL_{pi}$ formed by evaluating $W_{pi}(t)$ at every $t_{pi\ell}$. Furthermore, combine $\mbW_{pi}$ for all responses to form one vector $\mbW_{i}$ of length $\mrL_i$; $ \mbU_i$ and $\bPsi_i$ are defined analogously. Then $\mbW_i$ has mean  $\mrE(\mbW_i | \ba_i)= \mbU_i \bb + \bPsi_i \ba_i$ and precision matrix $\mb{P}_i$ is comprised of the appropriate error variance parameter $\tau_p^{-2}$.

 MCMC begins by setting initial values for all parameters and then sequentially sampling each parameter conditioned on all the others (denoted by `$` |\cdot$"). Sampling is performed (using the latest sample to update each parameter) according to the full conditionals in the following manner:
\begin{enumerate}[1.]
	\item Select initial values for $\bb$ , $\ba_i$, $\bSig$, $W_{pi}(t)$ for binary responses, and $\tau^2_p$ for continuous responses;
	\item  For each $i=1,\hdots,\mrN$ and $\ell=1,\hdots,\mrL_{pi}$, update the latent response corresponding to the observed binary response by drawing from  $W_{p i}(t_{i\ell})|\cdot \sim  \mrN(\mbu^\mrT_{pi}(t_{i\ell}) \bb + \bpsi^\mrT_p(t_{i\ell}) \ba_i, 1)$ restricted to the interval $(0,\infty)$ if $Y_{pi}(t_{i\ell})=1$ or $(-\infty,0)$ if $Y_{p i}(t_{i\ell})=0$;
	
	\item Update the population mean parameter by drawing from 
	$\bb | \cdot \sim \mrN( \mb{\mu}_{\bb}, \mbV_{\bb} )$ for
	  $\mbV_{\bb}= \left[ \left(\sum_{i=1}^\mrN \mbU_i^\mrT \mb{P}_i  \mbU_i \right)+  \sigma^{-2}_{\bb}\bI_\mrJ  \right]^{-1}$ and 
	  $\mb{\mu}_{\bb}  =  \mbV_{\bb} \left[ \sum_{i=1}^\mrN \mbU_i^\mrT \mb{P}_i  (\mbW_i - \bPsi_i \ba_i )\right]$;	
	
	\item For each  $i=1,\hdots,\mrN$, update the random effect by sampling from $\ba_i | \cdot \sim \mrN( \mb{\mu}_{\ba}, \mbV_{\ba} )$ for 
	$\mbV_{\ba}= \left( \bPsi_i^\mrT \mb{P}_i  \bPsi_i+ \bSig^{-1} \right) ^{-1}$ and
 $\mb{\mu}_{\ba} = \mbV_{\ba} \bPsi_i^\mrT \mb{P}_i (\mbW_i-\mbU_i \bb)$;

	\item Update the random effect covariance matrix through $\bSig| \cdot \sim \mr{InvWishart}_{\mrM}[ \{ \sum_{i=1}^\mrN{\ba_i\ba_i^\mrT}  +(1/q_2)\bI_{\mrM}\}^{-1}, \mrN+ q_1]$;
	
	\item Update the error variance for the continuous responses according to 
	$\tau^2_p | \cdot \sim \mr{InvGamma}(l_\omega, h_\omega)$ with $l_\omega =n/2+l$ and $h_\omega= h + 1/2 \sum_{i=1}^\mrN \sum_{\ell=1}^{{\mrL_i}} [ W_{p i}(t_{i\ell}) -  \mbu^\mrT_{pi}(t_{i\ell}) \bb + \bpsi^\mrT_p(t_{i\ell}) \ba_i ]^2$.

\end{enumerate}
Steps 2-6 are repeated for the desired number of samples.

%%%%%%%%%%%%%%%%%%%%%%%%%%%%%%
\section{Simulations}\label{s:sim}
%%%%%%%%%%%%%%%%%%%%%%%%%%%%%%

For our simulation study, we compare mean estimation and prediction performance among four estimating models when the generating model has a continuous and binary response, either generated separately (univariate) or jointly (bivariate) with strong cross-correlation, for both small and large sample size. The four estimating models are either univariate models applied separately to each response or bivariate models, and either employ the pre-specified basis function method of Section \ref{s:pre} or the data-driven approach of Section \ref{s:datadriven}.

	%%%%%%%%%%%%%%%%%%%%%
	\subsection{Data generation}\label{s:sim_data}
	%-----------------------------------------------
 We consider the case where $Y_{1i}(t)$ is continuous and $Y_{2i}(t)$ is binary. Functions are observed at a dense, balanced design with $\mrL_{pi} \equiv 30$ equally-spaced locations in $[0,1]$ for each subject $i$ and response $p$. We use the model given in \eqref{mod1} with predetermined bases as in Section \ref{s:pre} for data generation. We specify a separable random effect covariance matrix $\bSig= \mb{A} \otimes \mbC$, where $\Cov([\alpha_{1 ik}, \alpha_{2 ik}]^\mrT)=\mb{A}$ for $\mb{A}_{11}=\mb{A}_{22}=1$ and  $\mb{A}_{12}=\mb{A}_{21}=\rhoa$ so that the parameter $\rhoa$ controls the correlation between the latent responses, and $\Cov( [ \alpha_{p i1},\hdots, \alpha_{p i\mrM}]^\mrT)=\mbC$ for $p = 1, 2 $ controls the covariance of the random effect basis function coefficients and is the same across responses.  The $\mbC$ used for data generation has the AR(1) structure with variance 1 and correlation parameter $\rho=1/2$. 

For the fixed population mean function we assume there are no subject-level covariates so that $\bm{\mu}(t)=\mbB(t)\bb$, and we specify a quadratic basis $\{ \mrB_{p j}(t)=t^{(j-1)}: 1 \le j \le 3 \}$ for each response $p$ with coefficients $\bb_1 = [-0.64 , 4 , -4 ]^\mrT$ and $\bb_2=[0.97 ,   -6,    6]^\mrT$. The intercepts are chosen such that the curves are positive for approximately half of the observed locations $t$. The basis functions for the subject-specific deviation function $\mbf_i(t)= \bPsi(t) \ba_i$ are given by
$\psi_{1 k}(t)= \sin\{(2\pi k/ \mrM)(t+2\pi k/\mrM)\}$ and $\psi_{2 k}(t)= \cos\{(2\pi k/ \mrM)(t+2\pi k/\mrM)\}$ for $k=1,\hdots,\mrM=7$. The error variance for the continuous process is $\tau^2_1=1$.
We generate data from four scenarios given in Table \ref{t:sim2} by varying the sample size ($\mrN=50, 250$) and the cross-correlation ($\rhoa=0, 0.8$).  
All scenarios use $100$ Monte Carlo (MC) replications.

	%%%%%%%%%%%%%%%%%%%%%
	\subsection{Models and metrics for comparison}\label{s:sim_metrics}
	%-----------------------------------------------
	We fit four models to each dataset. 
	\begin{enumerate}[I.]%[itemsep=2pt,parsep=2pt]
		\item Bivariate B-spline (\BB): the multivariate model in \eqref{mod1} with B-spline bases as in Section \ref{s:pre};
		\item Univariate B-spline (\UB): the model from \eqref{latent1} applied separately to each response with B-spline bases as in Section \ref{s:pre};
		\item Bivariate FPCA (\BF): the multivariate model in \eqref{mod1} with data-driven bases as in Section \ref{s:datadriven};
		\item Univariate FPCA (\UF): the model in \eqref{latent1}  applied separately to each response with data-driven bases as in Section \ref{s:datadriven};
	
	\end{enumerate}
	For estimation using the B-spline methods, we choose B-splines of order 4 and the number of B-spline breaks for each replication is fixed at $6$ based on preliminary analyses.  For the FPCA methods, we specify an unstructured $\bSig$, and the number of basis functions is chosen to explain at least $\mr{P}_1=99\%$ of the cumulative variation. In practice, both the number of basis functions for the B-spline method and the percentage of variation explained for the FPCA method are tuning parameters and one should compare results over a grid parameter values. For the population mean we fit the true polynomial basis $\mbB(t)$ for estimation. We perform MCMC sampling with 20,000 draws and the first 5,000 are discarded as burn-in. The hyperparameters are specified as $\sigma^2_b=100$ and $q_1=q_2=l=h=0.1$.

Methods are compared in terms of their predictive performance and ability to estimate the marginal mean function for each response. Let $\omega_{1i}(t) = \mrE\{Y_{1 i}(t)\} = \mbu^\mrT_{1i}(t) \bb_1$ and $\omega_{2i}(t) = \mrE\{Y_{2 i}(t)\} =   \Phi \{ \gamma_i(t) \}$, where $\gamma_i(t) = \mbu^\mrT_{2i}(t)  \bb_2/\sqrt{ v_2(t) }$ is the population effect shrunk toward zero by the square root of the marginal variance 
$v_2(t) =\Var\{Y_2(t)\} = \bpsi_2(t) \bSig_{22} \{\bpsi_2(t)\}^\mrT + 1$.  
Let $\widehat{\omega}_{p r}(t)$ and $\hat{\nu}_{pr}(t)$ be the posterior mean and variance, respectively, for MC replication $r=1,\hdots,100$. 
 Metrics for comparison of estimated means found in Table \ref{t:sim2} for each response are mean integrated squared error: $\mr{MISE} = \int_t  \mrE \{ \widehat{\omega}_{p}(t) - {\omega}_{p}(t)   \}^2 dt$; coverage of 95\% pointwise confidence intervals $\widehat{\omega}_{pr}(t) \pm l_{pr}(t)$ averaged over location $t$ and MC replication $r$ with margin of error $l_{pr}(t)=1.96\sqrt{\hat{\nu}_{pr}(t)}$; and confidence interval length $2 l_{pr}(t)$. 
 
For prediction, we generate additional data $Y_{prj}(t_l)$ at equally spaced locations $t_\ell \in [0,1]$ where $\ell=1,\hdots,30$ for subjects $j=1,\hdots,20$ per response $p=1,2$  for each MC replication $r=1,\hdots,100$. To assess the value of jointly modeling the two responses, we leave out all of response 1 for 10 subjects and all of response 2 for the remaining 10 subjects per replication. 
%The response of interest is fitted for each subject within the MCMC algorithm described in Section \ref{s:compdetails}. For predicting the continuous response, one must add an additional step to the algorithm in which one draws from $W_{1jr}(t_{\ell})|\cdot \sim  \mrN(\mbU_1(t_{\ell}) \bb + \bpsi_1(t_{\ell}) \ba_{rj}, \tau^2_1)$. If predicting the binary response, replace step 2 in the algorithm by drawing from $W_{2 }(t_{\ell})|\cdot \sim  \mrN(\mbU_2(t_{\ell}) \bb + \bpsi_2(t_{\ell}) \ba_{rj}, 1)$ which is no longer restricted to a certain interval. Then the predicted response $\hat{Y}_{1rj}(t_\ell)$ is the posterior mean of $\mbU_1(t_{\ell}) \bb + \bpsi_1(t_{\ell}) \ba_{rj}$, and $\hat{Y}_{2rj}(t_\ell)$ is the posterior mean of $\Phi \big ( \mbU_1(t_{\ell}) \bb + \bpsi_1(t_{\ell})\ba_{rj} \big )$. 
Models are compared in terms of their predictive performance using mean squared prediction error (MSPE) for each response, defined as  $\mr{MSPE} =  (nm\mrL)^{-1} \sum_{r=1}^{n} \sum_{j=1}^{m} \sum_{\ell=1}^{\mrL}\{ Y_{prj}(t_\ell) - \hat{Y}_{prj}(t_\ell) \}^2$. For binary responses this is known as the Brier score and $\hat{Y}_{prj}(t_\ell)$ is the posterior probability that $Y=1$.

	%%%%%%%%%%%%%%%%%%%%%
	\subsection{Results}\label{s:sim_data}
	%-----------------------------------------------

Table \ref{t:sim2} gives the simulation results. There appears to be little difference in mean function estimation between univariate and bivariate methods for all scenarios. When strong correlation is present (Scenarios 1 \& 2), the bivariate methods show marked improvement in prediction for both responses over the univariate methods, a difference that becomes more pronounced with an increase in sample size. 
Bivariate methods perform well when the generating model is univariate (Scenarios 3 \& 4). Though prediction is better when fitting the correct univariate model, the differences between the bivariate and univariate methods become very small with an increase in sample size. All methods show slight under-coverage.

%------------------------------------------------------------------------------------------------------
\begin{table}
%\small
%\singlespacing
\caption{Simulation Results}
\centering
\resizebox{\linewidth}{!}
{%
	\begin{tabular}{l     llll  l   llll }
%	\toprule
\multicolumn{1}{c}{     }  & \multicolumn{4}{c}{ {Continuous Response}} & 
\multicolumn{1}{c}{     }   & \multicolumn{4}{c}{{Binary Response}} \\
 \cline{2-5}  \cline{7-10} \\

& MISE    & CI length & 95 \% Cvg  & MSPE & \vspace{.2cm}  & MISE   & CI length &  95 \% Cvg & MSPE   \\ 
\midrule
	 \multicolumn{10}{c}{Scenario 1: $n= 50$,  $ \rhoa=  0.8$ }  \\
\midrule 
BFPCA & 3.50        & 65.3         & 92.9 *** & 319    &     & 0.174      & 12.8         & 86.9 *** & 23.0          \\
BBSP  & 3.26       & 61.6 **      & 90.7 *** & 313       &  & 0.168      & 12.9         & 87.3 *** & 22.9        \\
UFPCA & 3.20        & 66.5 *       & 93.6     & 350 *    &   & 0.182      & 14.8 *       & 90.8 *** & 24.4 *      \\
UBSP  & 2.86       & 64.9         & 94.0       & 351 *     &  & 0.185      & 14.5 *       & 90.9 *** & 24.3 *      \\

\midrule
	 \multicolumn{10}{c}{Scenario 2: $n= 250$,  $ \rhoa=  0.8$ }  \\
\midrule 
BFPCA & 0.795      & 31.9         & 91.4 *** & 284     &    & 0.039     & 6.66         & 90.7 *** & 21.0          \\
BBSP  & 0.798      & 31.2 **      & 91.0 ***   & 285   &      & 0.037     & 6.67         & 91.3 *** & 21.0          \\
UFPCA & 0.790      & 32.8 *       & 91.9 *** & 351 *  &     & 0.040     & 7.27 *       & 93.2     & 24.3 *      \\
UBSP  & 0.794      & 32.4 *       & 92.0 ***   & 350 * &      & 0.043     & 7.25 *       & 91.6 *** & 24.3 *      \\

\midrule
\midrule
	 \multicolumn{10}{c}{Scenario 3: $n= 50$,  $ \rhoa=  0$ }  \\
\midrule 
BFPCA & 2.96       & 65.9         & 94.2     & 408       &  & 0.172      & 13.4         & 89.3 *** & 26.3        \\
BBSP  & 3.16       & 62.6 **      & 92.8 *** & 421 *    &   & 0.183      & 13.3         & 88.4 *** & 26.6 *      \\
UFPCA & 2.75       & 65.8         & 94.6     & 372 **   &   & 0.166      & 14.6 *       & 93.3     & 24.4 **     \\
UBSP  & 2.85       & 63.9 **      & 93.5     & 371 **   &   & 0.162      & 14.5 *       & 92.8 *** & 24.2 **     \\

\midrule
	 \multicolumn{10}{c}{Scenario 4: $n= 250$,  $ \rhoa=  0$ }  \\
\midrule 
BFPCA & 0.802      & 32.5         & 94.6   & 370        & & 0.044     & 6.96         & 91.1 *** & 24.8        \\
BBSP  & 0.791      & 31.9 **      & 94.3   & 374 *      & & 0.042     & 6.97         & 90.7 *** & 24.9        \\
UFPCA & 0.780       & 32.9 *       & 94.7   & 362 **   &   & 0.042     & 7.35 *       & 93.5     & 24.3 **     \\
UBSP  & 0.765      & 32.5         & 94.5   & 361 **     & & 0.040     & 7.35 *       & 94.1     & 24.3 **     \\

\midrule
\multicolumn{10}{p{\linewidth}}{{Results in hundredths. A `**' (`*') indicates better (worse) compared to BFPCA by Wilcoxson rank sum test, $\alpha=0.05$. For coverage, a `***' indicates that the coverage is not within the nominal $95\%$ range. }}\\
	\bottomrule
	\end{tabular}
}
\label{t:sim2}
\end{table} 
%------------------------------------------------------------------------------------------------------

For Scenarios 1 \& 2 there is no clear difference between fitting predetermined bases (Section \ref{s:pre}) or data-driven bases (Section \ref{s:datadriven}); however, BFPCA has better prediction compared to BBSP in Scenarios 3 \& 4 when there is no cross-correlation. The univariate models have very similar performance to one another in all scenarios.

	%%%%%%%%%%%%%%%%%%%%%
	\section{Periodontal Data Application}\label{s:dental_data}
	%-----------------------------------------------

We demonstrate our methods using data from a periodontal study \citep{Fer+:09} conducted by the Center for Oral Health
Research at the Medical University of South Carolina. In addition to collecting subject-level covariates for over 200 Gullah African Americans, several measures of patients' periodontal health were observed at six sites for each of 28 teeth. The two responses we consider are (continuous) clinical attachment loss (CAL) and (binary-valued) bleeding on probing (BOP). CAL is the distance that a tooth has detached from the bone, rounded to the nearest mm. We use the average CAL over the six sites on each tooth as the tooth's CAL response. 
BOP is the binary indicator of whether the gums bleed when pressed with a dental probe at any of the six sites per tooth. A total of $\mrN=197$ patients (subjects) are included for analysis after excluding those with more than 50\% missingness. Any remaining missingness is assumed to be completely at random; \cite{Rei+Ban:10} and \cite{Rei+:13} provide methods for accounting for non-random missingness.

For our analysis, we assign teeth the numbers 1-14 going from left to right in the upper jaw when looking at a patient and 15-28 going from right to left in the lower jaw when looking at a patient; wisdom teeth are excluded. Using this numbering system, teeth 1 \& 28  are adjacent going from upper jaw to lower jaw, and it is the same for teeth 14 \& 15 on the other side of the mouth. 
We consider responses at each tooth to be realizations of a functional process with locations $t \in [1,28]$. In fitting a bivariate functional model to this data, we hope to gain a better understanding of the dynamics between the responses CAL and BOP through close examination of their cross-covariance. Our extremely flexible approach to modeling the covariance will be able to capture any spatial correlation of adjacent teeth, of teeth on different sides of the mouth, and of teeth on different jaws.

The subject-specific covariates  that we include in modeling the mean function are the same covariates used by \cite{Rei+Ban:10} and include age (in years), gender (female=1, male=0), body mass index or BMI (in $\mr{kg}/\mr{m}^2$), smoking status (1=smoker, 0=never), and glycosylated hemoglobin or HbA1c (1 = high, 0 = controlled).  All covariates have been standardized to be zero-mean with standard deviation of 1. For each tooth, we include an indicator of jaw (0=upper, 1=lower). For the smooth part of the mean, we consider a quadratic function 
$s_p(t) = \beta_{p0} + \beta_{p1}d(t) + \beta_{p2}d(t)^2 $ of tooth distance $d$ from the front of the mouth, where $d(t)= t - 7.5$ for teeth in the upper jaw and $d(t)= t- 21.5$ for the lower jaw.

We present analysis for 8 models given in Table \ref{t:data2} that all employ the data-driven basis method of Section \ref{s:datadriven}. 
The 8 models differ by: 1) whether FPCA is univariate or bivariate; 2) the choice of threshold $\mrP_1=99\%,95\%$ for the cumulative percentage of variation explained for FPCA; and 3) whether a random bivariate subject-level intercept $\ba_{0i}= [\alpha_{01i}, \alpha_{02i}]^\mrT$ is added to model \eqref{mod1}. Models using B-splines as in Section \ref{s:pre} were also considered but are not presented because the best-performing models required a large number of basis functions.

%------------------------------------------------------------------------------------------------------
\begin{table}
\normalsize
%\singlespacing
\caption{Model comparisons for the periodontal data application}
\centering
\resizebox{\linewidth}{!}
{%
	\begin{tabular}{rrrrrrrr }
Model & Subject RE &   PVE & PCA &Dbar & pD & DIC \\
\midrule 
1 & Y & 99 & B    & 8114 & 1257 & 9371 \\
2 & Y & 99 & U    & 8228 & 1231 & 9459 \\
3 & Y & 95 & B    & 8849 & 1001  & 9850  \\
4 & Y & 95 & U    & 8536 & 1114 & 9650 \\
5 & N & 99 & B    & 10101 & 942 & 11043 \\
6 & N & 99 & U    & 10586 & 832 & 11418 \\
7 &  N& 95 & B    & 10511 & 788 & 11299 \\
8 & N & 95 & U    & 10722 & 760 & 11482 \\ 

\midrule
\multicolumn{7}{p{\linewidth}}{\small ``Subject RE" indicates inclusion of a subject-specific random intercept. ``PVE" is the threshold for cumulative percentage of variation explained. ``PCA" indicates whether univariate (``U") or bivariate (``B") FPCA was performed. }\\
\bottomrule
\end{tabular}
}
\label{t:data2}
\end{table} 
%------------------------------------------------------------------------------------------------------

For the purpose of estimating the latent covariance, we ignore the covariates.   When incorporating a bivariate random subject-level intercept, we use residuals $R_{1i}(t)= Y_{1i}(t) - \mrL_{1i}^{-1}\sum_{i=1}^{\mrL_{1i}} Y_{1i}(t)$ of the continuous response CAL to estimate the latent covariance for FPCA; this is not done for the binary responses as the residuals would no longer be binary. For models that include $\ba_{0i}$, we estimate the covariance term $\Cov(\alpha_{01i}, \alpha_{02i})$ in addition to the variance terms $\Var(\alpha_{0pi})$. We specify a diagonal covariance matrix $\bSig$ for the remaining random effect parameters.

\begin{figure}
\centering
%\captionsetup{justification=raggedright,margin=1cm}
%\includegraphics[width=.85\textwidth]{plots/bxplt_1}
\includegraphics[width=\linewidth]{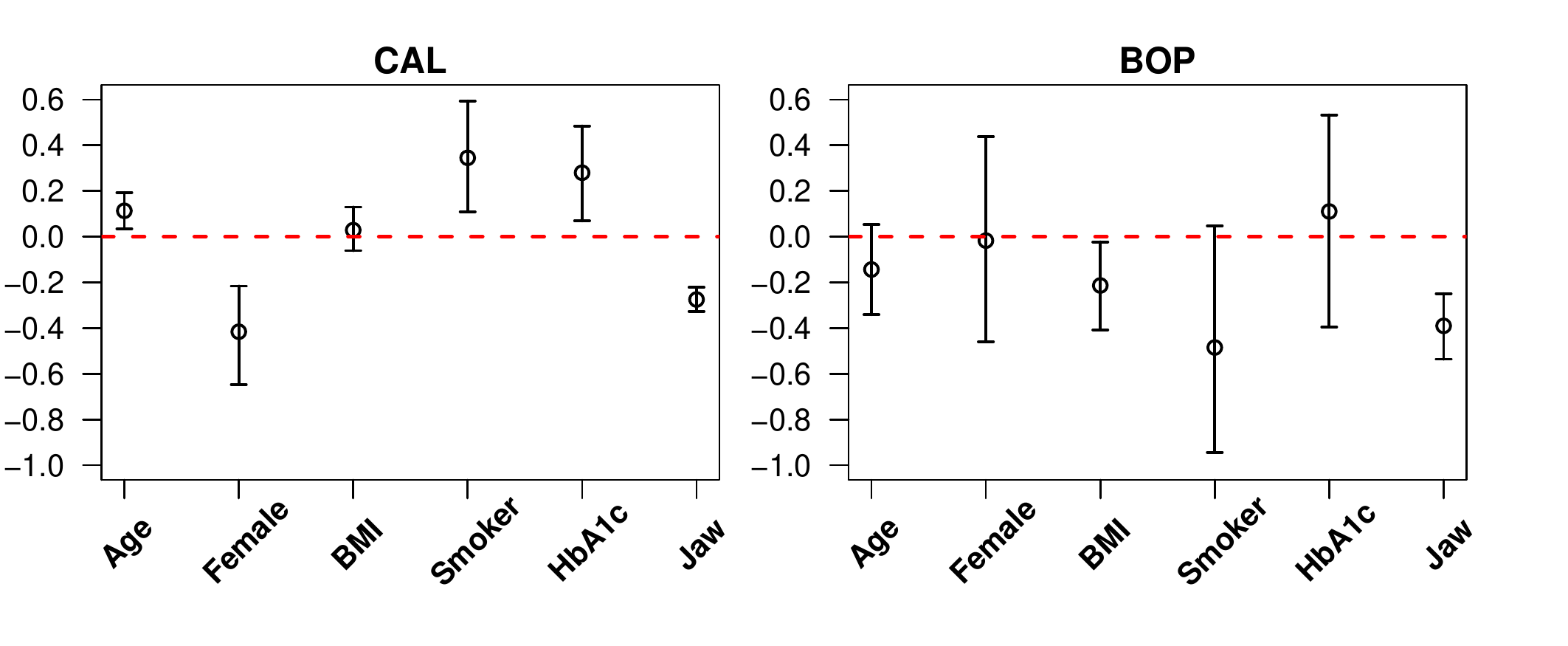}
\caption{Posterior medians and 95\% posterior intervals of the subject-specific covariate coefficients by response.}
\label{f:covariates}
\end{figure}

Table \ref{t:data2} shows that Models 1-4, which include a subject random effect, outperform (based on DIC) Models 5-8 which omit the subject random effect. For this data, specifying the larger percentage of variation explained for FPCA, and hence including more basis functions, leads to better model performance. In comparing the two leading models 1 \& 2, implementing FPCA on the full bivariate covariance matrix as in Model 1, taking into account the cross-dependence between the two responses CAL and BOP, leads to superior performance.
Figure \ref{f:covariates} shows the subject-level coefficient estimates and 95\% posterior intervals for Model 1. Models 2-4 had similar coefficient estimates.  For CAL, only the coefficient interval for BMI includes zero. The other coefficient estimates show an increased level of CAL for older patients, males, smokers, patients with high HbA1c counts, and for teeth on the upper jaw. For BOP, the posterior confidence intervals are larger than those for CAL. For intervals that exclude zero, there is an increase of BOP for the upper jaw, yet a slightly lower incidence of BOP for higher BMI.

%------------------------------------------------------------------------------------------------------
%Model & Response & Age  & Gender & BMI & Smoking  & HbA1c & Jaw \\ 
%Estimate & 1 & CAL & 0.11* & -0.41* & 0.03 & 0.34* & 0.28* & -0.27* \\
% SD       & 1 & CAL & 0.05  & 0.13   & 0.06 & 0.15  & 0.12  & 0.03   \\                                                        
% Estimate & 1 & BOP & -0.14 & -0.02 & -0.21* & -0.49* & 0.13 & -0.39* \\
% SD       & 1 & BOP & 0.12  & 0.27  & 0.12   & 0.3    & 0.28 & 0.09   \\
 %------------------------------------------------------------------------------------------------------

\begin{figure}
\centering
%\captionsetup{justification=raggedright,margin=1cm}
\includegraphics[width=\linewidth]{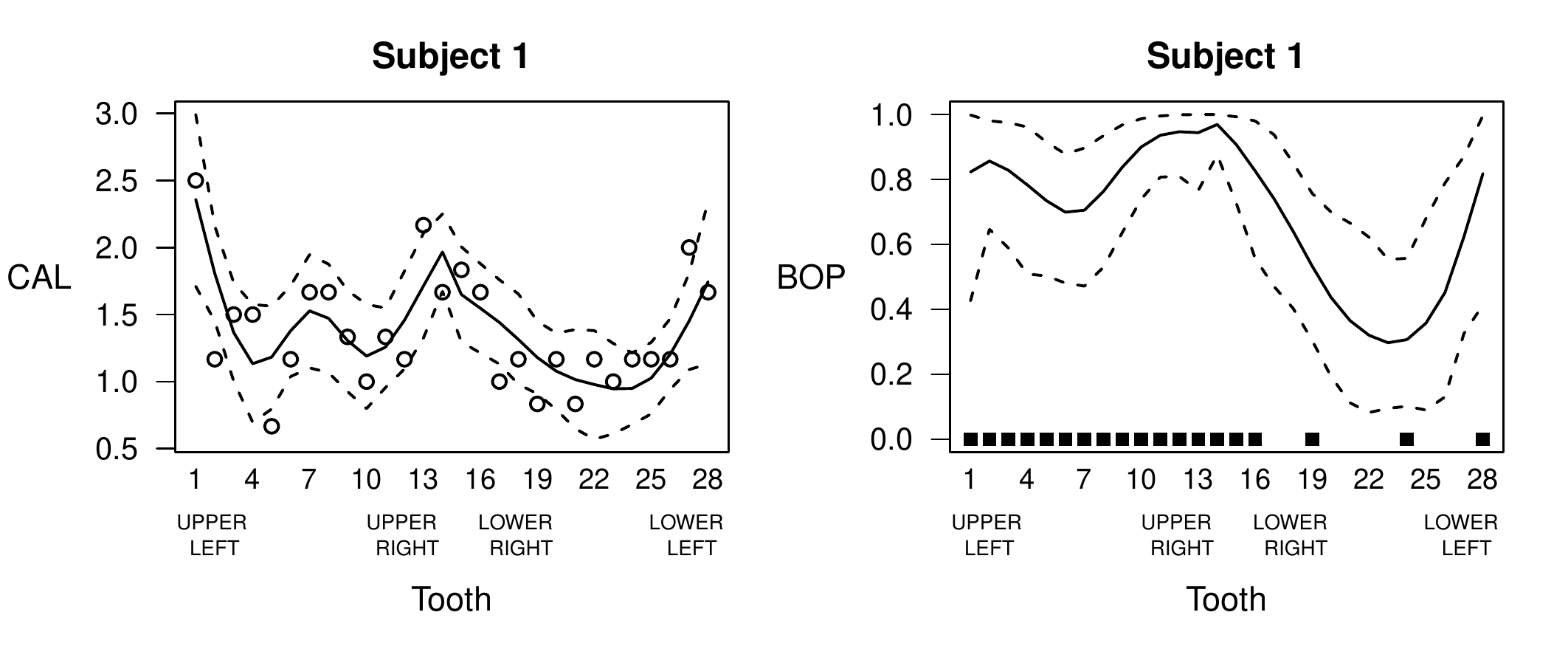}
\includegraphics[width=\linewidth]{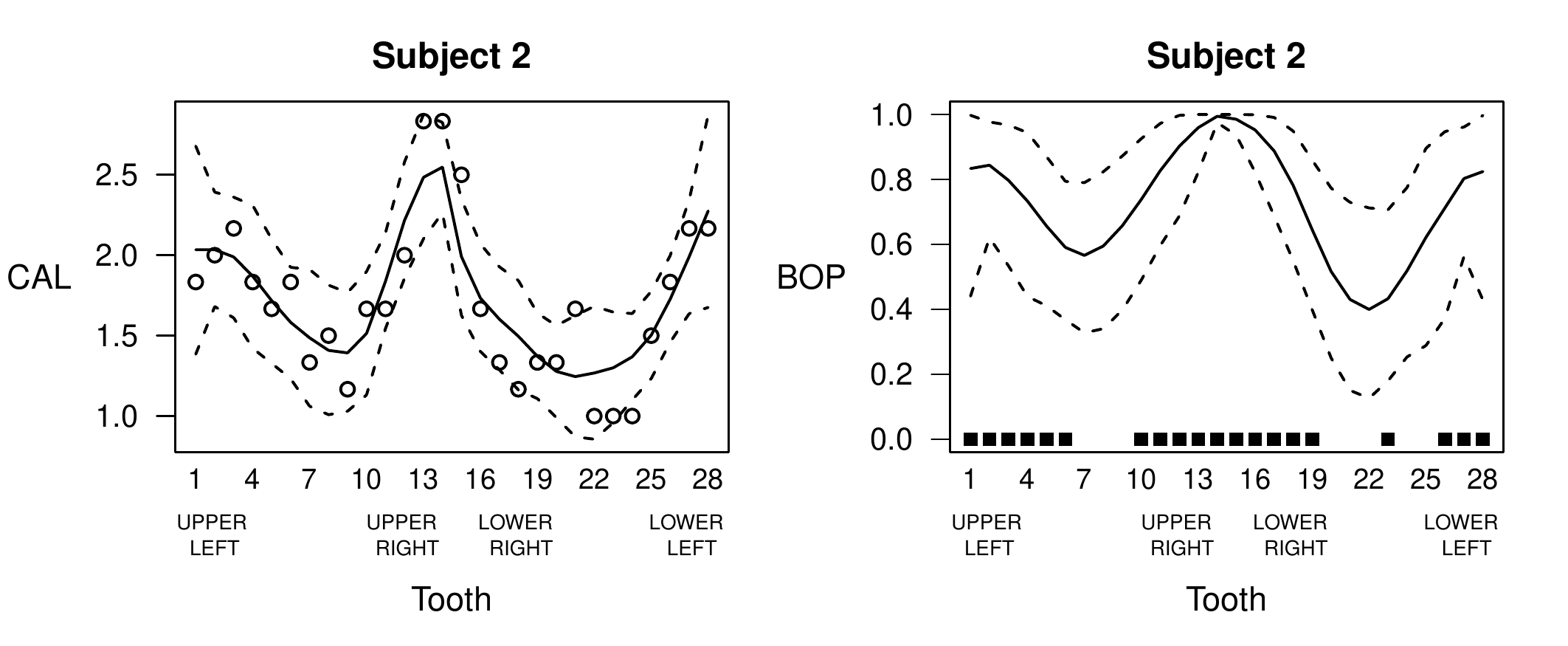}
\caption[Fitted values for two individuals from the periodontal study.]{Fitted values for two individuals from the periodontal study (using Model 1). \emph{Left panels}: Observed values of CAL are shown as dots. The solid black line indicates the posterior mean of $\mu_{1i}(t)$, the subject-specific mean function, and point-wise 95\% posterior intervals are given by the dotted lines. \emph{Right panels}: The squares along the x-axis indicate the teeth for which BOP is observed. The solid black line gives the posterior mean of the conditional probability of the event, $P(Y_{2i}(t)=1|\ba_{2i})$, and dotted lines show point-wise 95\% posterior intervals. The label ``UPPER LEFT" refers to the left side of the the upper jaw when looking at a patient, and it is analogous for the other labels.}
\label{f:fitted}
\end{figure}

Figure \ref{f:fitted} shows the fitted values (from Model 1) for two individuals in the periodontal data set. The left panels show the posterior means and 95\% posterior intervals of the subject-specific mean function $\mu_{1i}(t)$ for the continuous response CAL. Most of the observed CAL values fall within the 95\% interval for both subjects, indicating a reasonable model fit. The right panels show the posterior mean and 95\% posterior intervals of the conditional probability of the event, $P(Y_{2i}(t)=1|\ba_{2i})$. Teeth with observed BOP ($=1$) are indicated by the squares on the bottom of the plot. The higher predicted probabilities tend to correspond to the incidence of BOP, again indicating a reasonable model fit.

%------------------------------------------------------------------------------------------------------
%------------------------------------------------------------------------------------------------------

\begin{figure}
\centering
%\captionsetup{justification=raggedright,margin=1cm}
%\includegraphics[width=.85\textwidth]{plots/bxplt_1}
\includegraphics[width=\linewidth]{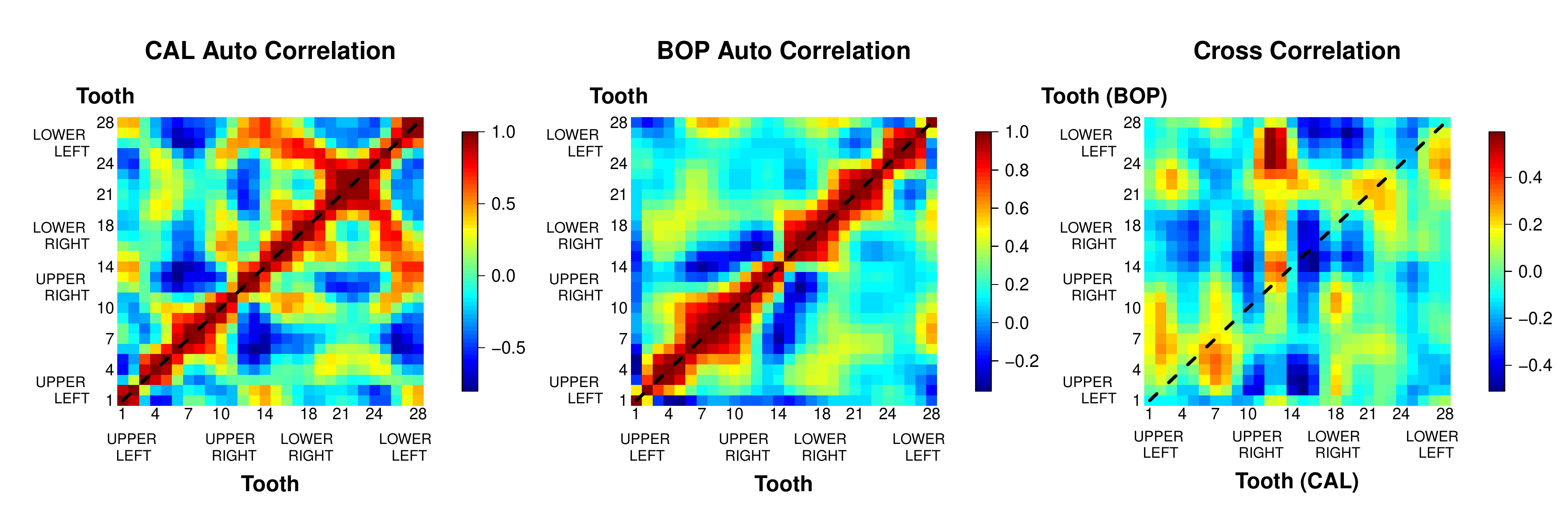}
\includegraphics[width=\linewidth]{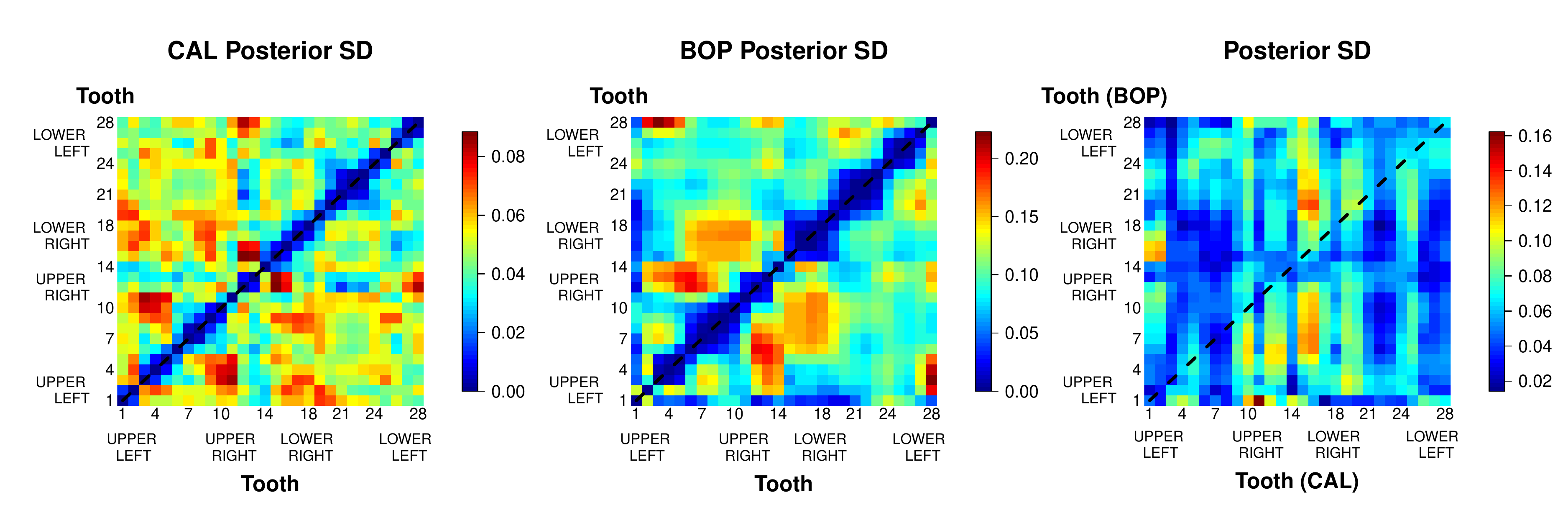}
\includegraphics[width=\linewidth]{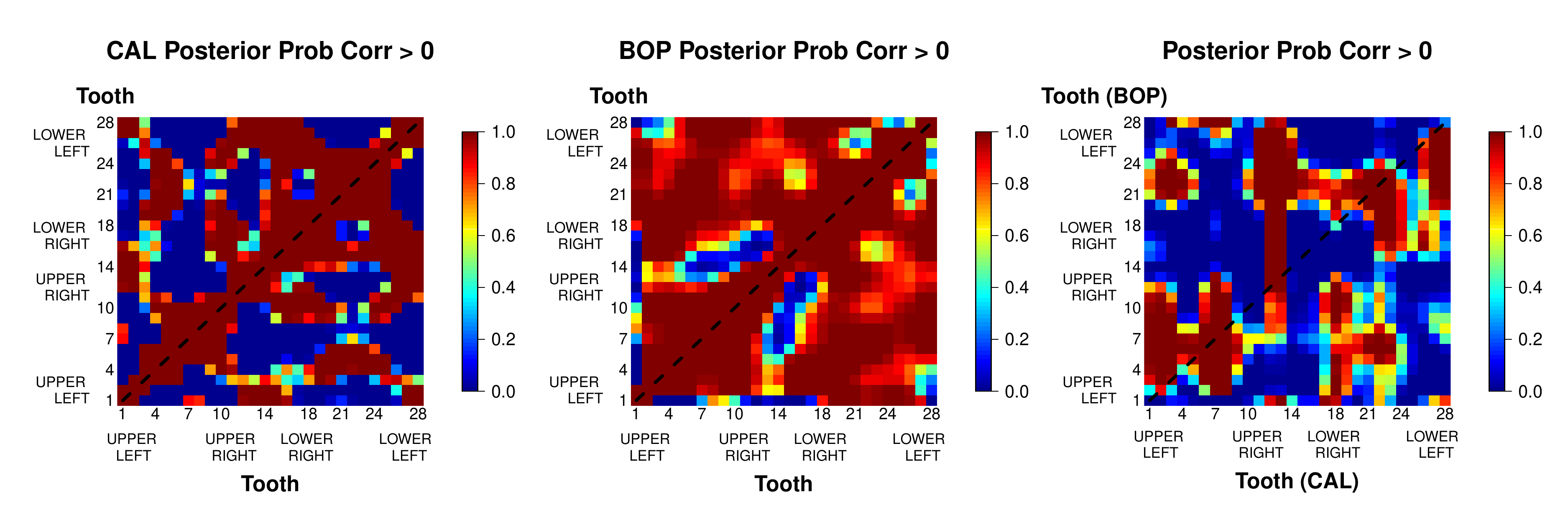}
\caption{Posterior summaries of the within-response and between-response correlation structures for any two teeth when fit with Model 1 (excluding correlation from the subject random intercepts). The label ``UPPER LEFT" refers to the left side of the the upper jaw when looking at a patient, and it is analogous for the other labels.}
\label{f:post_corr}
\end{figure}

The posterior summaries of the auto- and cross-correlations of the subject-specific process $\mb{f}_i(t)$ from \eqref{mod1} are given in Figure \ref{f:post_corr}; note that the correlation attributed to the subject random intercept is not included in this figure. In this periodontal application, these plots offer important and novel insights into the complex relationships that exist between and within the BOP and CAL responses in different parts of the mouth. The utility of quantifying and visualizing these complex correlation relationships is apparent for many other types of applications.

%------------------------------------------------------------------------------------------------------
%------------------------------------------------------------------------------------------------------

Examination of the diagonal of the auto-correlation plot for CAL in Figure \ref{f:post_corr} shows strong positive spatial correlation between adjacent teeth and between teeth separated by only one or two teeth on the same jaw.  This plot also shows positive correlation between a tooth in the left and a tooth in the right side of the same jaw, and the relationship is particularly strong for teeth in the lower jaw. The correlation for CAL between teeth in opposite sides of the mouth and on different jaws is also positive, yet not as strong as for teeth on the same jaw; this correlation is very similar in magnitude as the correlation for teeth on the left or right side of the mouth but on different jaws. Additionally, there are mild to strong negative correlations between teeth in the center (front) of the mouth and teeth in the back of the mouth, regardless of the jaws on which the teeth are located. This is also seen in the plot of the posterior probability that the auto-correlation is positive.

In the auto-correlation plot for BOP, again we see strong positive spatial correlation between adjacent teeth and between teeth that are close to one another on the same jaw. Additionally, the plots of the auto-correlation and of the probability of being positive show that the correlation is mostly positive with only a few areas of negative correlation. The correlation is negative between a tooth in the center and a tooth on the right side of the upper jaw, as well as between a tooth in the left and a tooth in the center of the lower jaw. There is also a strong negative correlation for teeth in the lower right and upper right, as well as for teeth in the lower left and lower right. 

The cross-correlation between BOP and CAL ranges from moderately positive to moderately negative. Unlike the auto correlation plots, the cross correlation is not symmetric, which makes interpretation slightly more complex. For instance, BOP in the lower left is positively correlated with CAL in the center and upper right as indicated by the darkest patch near the top center of the cross correlation figure. Alternatively, CAL in the lower left shows slightly negative to no correlation with BOP in the center and upper right of the mouth.  
Another demonstration of this non-symmetric property occurs for the negative correlation of BOP in the lower left with CAL in the lower right, though BOP in the lower right shows slightly positive to no correlation with CAL in the lower left.

%-------------------------
\section{Discussion}
%--------------------------

We introduce a methodology to jointly model multivariate functional responses of mixed type (e.g. continuous and binary data) and also propose an extension of FPCA for mixed-responses. 
Our method can account for subject-specific covariates that can be either linear or time-dependent (such as the jaw indicator used in the analysis of the periodontal study in Section \ref{s:dental_data}). 
The proposed method is flexible enough for functions to be observed at varying locations for different subjects and different responses.
For exposition we focus on modeling a bivariate response vector where one functional response is continuous and the other is binary, though joint modeling of more than two responses is a straightforward extension. Furthermore, the method easily models repeatedly observed categorical responses. This is achieved in a manner similar to thresholding the latent process at zero for binary data, but instead one must impose multiple thresholds on the latent process. Modeling other types of data, such as repeatedly observed count data, is not as straightforward as it would likely require using copulas% \citep{nelsen-1999}. 

By estimating the multivariate covariance of the latent process, our methodology can offer novel insights into the cross-dependence of different responses, which is of interest in a wide
variety of applications. Quantifying and exploring this dependence is an important contribution of our method and is a primary goal of our analysis of the periodontal data presented in Section \ref{s:dental_data}.  \cite{Rei+Ban:10} and \cite{Rei+:13} offer ways to incorporate informative missingness and apply their methods to the same periodontal data. We do not address the informative missingness for our analysis because it is not central to our goals, and leave it for future work.

%\backmatter

\section*{Acknowledgments}
The authors thank Dipankar Bandypadhyay of the University of Minnesota and Drs. S. London, J. Fernandes, C. Salinas, W. Zhao, Ms. L. Summerlin and Ms. P. Hudson. W of the Center for Oral Health Research (COHR) at the Medical University of South Carolina for providing the data and context for this work.  The work of Beth A. Tidemann-Miller was supported by NIH grant R01 CA085848.

\bibliographystyle{rss}
\bibliography{researchbib2}

\section*{Appendix}

\section*{Appendix A}

In this section, we show that we can approximate any smooth covariance using the predetermined basis method.  
For simplicity, assume that the functional responses are observed at the same locations $t_{p\ell} \equiv t_\ell$ for $\ell = 1, \hdots, \mrL$ for each response $p$. We specify this model for an arbitrary subject, and thus drop the subscript $i$. Let $\bpsi_{pk}$ be the vector  of length $\mrL$ formed by evaluating at every $t_{\ell}$ the basis functions $\psi_{pk}(t)$, $k=1,\hdots,\mrM_p$, and define the vector $\mbf_{p}$ analogously. Then we form the $\mrL \times \mrM_p$ matrix $\bPsi_{p}=[\bpsi_{pk}, \hdots, \bpsi_{p \mrM_p}]$ and the coefficient vector $\ba_{pi} = [\alpha_{p1},\hdots,\alpha_{p{\mrM_p}} ]^\mrT$ so that we can write $\mbf_{p}=\bPsi_{p} \ba_{p}$. We combine $\mbf_{p}$ for all responses to form one vector $\mbf$ of length $\mrn= \mrP \mrL$, and define the coefficient vector $\ba^\mrT = [\ba_{1}^\mrT,\hdots, \ba_{\mrP}^\mrT]$ of length $\mrm=\sum_{p=1}^\mrP \mrM_p$ and corresponding block-diagonal matrix $\bPsi$ of dimension $\mrn \times \mrm$ with blocks $\bPsi_{p}$. The resulting vector  $\mbf= \bPsi \ba$ has length $\mrn$, and we assume $\ba \overset{iid}{\sim}N(0,\bSig)$ where $\bSig$ is a covariance matrix of dimension $\mrm \times \mrm$ with elements $\Cov(\alpha_{pk},\alpha_{p\pr  \ell })= \xi_{k \ell p p\pr}$.

To illustrate the flexibility of the model, assume $\mb{\Omega}_0$ is the true $\mrn \times \mrn$ covariance matrix of $\mbf$ evaluated at locations $t_{l}$. $\mb{\Omega}_0$ is now approximated by the variance-covariance matrix 
$\mb{\Omega}=\Cov(\bPsi \ba) = \bPsi \bSig \bPsi^\mrT$. Since the basis comprising $\bPsi$ is pre-specified, the quality of the approximation $\mb{\Omega}\approx \mb{\Omega}_0$ is reliant on $\bSig$. By fitting a large number of basis functions, i.e. setting $\mrm=\mrn$, it is possible to fit any smooth covariance function. When $\mrm=\mrn$ then $\bPsi_i$ is a square matrix. Assume $\bPsi_i$ is full rank and thus $\bPsi_i^\mrT\bPsi_i $ is invertible. Pre- and post- multiplication gives $\bPsi_i^\mrT \mb{\Omega} \bPsi_i = \bPsi_i^\mrT \bPsi_i \bSig \bPsi_i^\mrT \bPsi_i$. Since $\mb{\Theta}=\{\bPsi_i^\mrT\bPsi_i\}^{-1}$ exists we can recover $\bSig = \mb{\Theta}\bPsi_i^\mrT \mb{\Omega} \bPsi_i \mb{\Theta}$. Though this approach is quite flexible, it is 
hard to estimate $\bSig$ if it is high-dimension; therefore it is unlikely to perform well if the processes cannot be represented by a small number of basis functions.

%%%%%%%%%%%%%
\section*{Appendix B}

Here we describe in more detail the derivation of the latent cross covariance estimator. As our approach is inspired by \cite{Hal+:08}, we start with a brief summary of the method they proposed for finding the auto-covariance of the latent process corresponding to the binary response, that is, response $p=2$. First, estimate the mean function for $p=2$, $\hat{\mu}_2(t)=g^{-1}\{ \hat{\eta}_2(t) \}$ where $\hat{\eta}_2(t)$ estimates $\mrE[\gzt ]={\eta}_2(t)$ and is found by smoothing the data $\big(t,Y_{2 i}(t)\big)$ for $i=1,\hdots,\mrN$. Next, find the estimator $\hat{S}_{22}(t,t\pr)$ of $S_{22}(t,t\pr)= \mrE\{Y_{2 i}(t)Y_{2 i}(t\pr)\}= \mrE\big[  g\{Z_{2i}(t)\}  g\{Z_{2i}(t\pr)\} \big]$ by performing bivariate smoothing of the data  $\big((t,t\pr),Y_{2 i}(t) Y_{2 i}(t\pr) \big)$ for $i=1,\hdots,\mrN$, once again removing the diagonals before smoothing. The estimator of the latent process covariance operator for the second response is given by 
\begin{align} \label{Hall_univ}
\widetilde{K}_{22}(t,t\pr) = \{\hat{S}_{22}(t,t\pr)- \hat{\eta}_2(t)\hat{\eta}_2(t\pr) \} /[ g^{(1)}\{\hat{\mu}_2(t)\}  g^{(1)}\{\hat{\mu}_2(t\pr)\}   ].
\end{align}

Equation \eqref{Hall_univ} was developed for a univariate response, so in order to estimate the latent cross covariance operator $K_{12}(t,t\pr) = K_{21}(t\pr,t)= \Cov\{Z_{1 i}(t), Z_{2 i}(t\pr)\}$, we must derive an analogous estimator. This requires the following Taylor expansion, also given by equation (5) in \cite{Hal+:08},
\begin{align}\label{Taylor1}
g\{Z_i(t)\}=&g\{\mu(t)\}+\delta X_i(t) g^{(1)}\{\mu(t)\}+ \frac{1}{2}\delta^2 \{X_i(t)\}^2 g^{(2)}\{\mu(t)\} \nonumber\\
&+\frac{1}{6}\delta^3 \{X_i(t)\}^3g^{(3)}\{\mu(t)\}+ O_p(\delta^4).
\end{align}
We can expand the covariance of the observed processes $ \Cov \big \{ Y_{1 i}(t),Y_{2 i}(t\pr) \big \}= \Cov \big[Z_{1 i}(t),g\{Z_{2 i}(t\pr)\} \big]$ by substituting \eqref{Taylor1} for $g\{Z_{2 i}(t\pr)\}$ and $\mu_1(t) + \delta X_{1 i}(t)$ for $Z_{1 i}(t)$, which gives
\begin{align} \label{Taylor}
\Cov \big\{Y_{1 i}(t),Y_{2 i}(t\pr) \big\} &=	g^{(1)}\{\mu_2(t\pr)\} \Cov\{\delta X_{1 i}(t), \delta X_{2 i}(t\pr) \}   +  O(\delta^4).
\end{align}	 
Note that the term (suppressed from equation \eqref{Taylor}) $\delta^3 \frac{1}{2}g^{(2)}\{\mu_2(t\pr)\} \Cov\{X_{1 i}(t),  X^2_{2 i}(t\pr) \}=0$ due to 
$\Cov\{X_{1 i}(t),  X^2_{2 i}(t\pr) \}  = \mrE\{X_{1 i}(t) X^2_{2 i}(t\pr) \} =
\mrE[X^2_{2 i}(t\pr) \mrE\{X_{1 i}(t) |X_{2 i}(t\pr) \}]=\sigma_1/\sigma_2\rho \mrE[  X^3_{2 i}(t\pr) ] = 0$ since $X_{1 i}(t) |X_{2 i}(t\pr) \sim \mrN\big(\sigma_1/\sigma_2\rho X_{2 i}(t\pr) ,(1-\rho^2)\sigma^2_1\big)$. 
Now, because $\Cov \{Z_{1 i}(t),Z_{2 i}(t\pr)\} = \Cov\{\delta X_{1 i}(t), \delta X_{2 i}(t\pr) \}$,  we have from \eqref{Taylor} that 
${K}_{12}(t,t\pr) =  \Cov \{Z_{1 i}(t),Z_{2 i}(t\pr)\}=\Cov \big \{ Y_{1 i}(t),Y_{2 i}(t\pr) \big \}/g^{(1)}\{\mu_2(t\pr)\} +O(\delta^4)$,
which, assuming the effect of $O(\delta^4)$ is negligible, leads to
\begin{align} \label{cross}
{K}_{12}(t,t\pr) = \Cov \big\{Y_{1 i}(t),Y_{2 i}(t\pr) \big \}/g^{(1)}\{\mu_2(t\pr)\}. 
\end{align}

Estimation of \eqref{cross} requires a smooth estimate $\hat{\eta}_1(t)$ of $\mrE[Y_{1 i}(t) ]={\eta}_1(t)$ which is found by smoothing the data $\big(t,Y_{1 i}(t)\big)$ for $i=1,\hdots,\mrN$. We obtain the estimator $\hat{S}_{12}(t,t\pr)$ of $S_{12}(t,t\pr)= \mrE\{Y_{1 i}(t)Y_{2 i}(t\pr)\}= \mrE\big[ Y_{1 i}(t) g\{ Z_{2i}(t\pr) \} \big]$ by performing bivariate smoothing of the data  $\big((t,t\pr),Y_{1 i}(t) Y_{2 i}(t\pr) \big)$ for $i=1,\hdots,\mrN$, removing the diagonals before smoothing. The resulting smooth estimator of the latent cross covariance is 
\begin{align} \label{cross2}
\widetilde{K}_{12}(t,t\pr) = \{\hat{S}_{12}(t,t\pr)- \hat{\eta}_1(t)\hat{\eta}_2(t\pr) \} / g^{(1)}\{\hat{\mu}_2(t\pr)\},
\end{align}
which is the direct analogue to \eqref{Hall_univ}. 

%%%%%%%%%%%%%
\section*{Appendix C}

In this section we present the derivations of the conditional posterior distributions. 

\subsection*{Random effects}
Let $\mrL_i$ be the number of subunits $t$ observed for subject $i$ and define the latent response vector for subject $i$ as $\mbW_i = [W_\c(t_1),...,W_\c(t_{\mrL_i}), W^2(t_1),...,W^2(t_{\mrL_i})]^\mrT$, with corresponding mean vector $\mrE(\mbW_i)=\mbU_i \bb + \bPsi_i \ba_i$.  Assume $\mbW_i|\ba_i \sim \mrN_{2 \mrL_i} \big(\mbU_i \bb + \bPsi_i \ba_i, \mb{D}_i\big)$ where $\mb{D}_i=\mr{diag}(\tau^2_\c,1) \otimes \mb{\mr{I}}_{\mrL_i} $, or in terms of the precision, $\mb{P}_i=\mb{D}_i^{-1}=\mr{diag}(\omega_\c, 1) \otimes \mb{\mr{I}}_{\mrL_i} $.

Also assume the $\mrm \times 1$ vector $\ba_i|\mbQ \sim {\mrN}_\mrm(0, \mbQinv)$ for $i=1,\hdots,\mrN$ and for the $\mrm \times \mrm$ covariance matrix $\mbQinv$, or equivalently, the precision matrix $\mbQ$. Define $\mbR_i=\mbW_i-\mbU_i \bb$. To find the posterior for $\ba_i | \cdot$ we know 
\begin{align}
p(\ba_i | \cdot) & \propto   p(\mbW_i | \cdot    )           \times p(\ba_i| \mbQ)   \\
& \propto  \exp \left[  -\frac{1}{2} \left \{
( \bPsi_i\ba_i -  \mbR_i )^\mrT \mb{P}_i ( \bPsi_i\ba_i -  \mbR_i )+ \ba_i^\mrT \mbQ \ba_i    
\right \}  \right ],  \\
& \propto  \exp \left[  -\frac{1}{2} \left \{
\ba^\mrT_i \bPsi_i^\mrT\mb{P}_i  \bPsi_i\ba_i 
-2 \mbR^\mrT_i \mb{P}_i  \bPsi_i\ba_i 
+ \mbR_i^\mrT\mb{P}_i\mbR_i
+  \ba_i^\mrT \mbQ \ba_i
\right \}  \right ],  \\			     		     
\intertext{\centering and ignoring terms not involving $\ba_i$ or $\mbQ$ results in}  
p(\ba_i | \cdot) & \propto  \exp \left(  -\frac{1}{2}  \left [  -2 \mbR^\mrT_i \mb{P}_i\bPsi_i\ba_i   +  
\ba_i^\mrT \left\{ \bPsi_i^\mrT \mb{P}_i  \bPsi_i+ \mbQ \right \}\ba_i       \right]  \right ). 
\end{align}
We want to form this term into the kernel of a Gaussian distribution  where the exponent is $ -1/2(\ba_i - \mb{M})^\mrT \mbV^{-1}  (\ba_i - \mb{M})= -1/2(\ba_i^\mrT \mbV^{-1} \ba_i  - 2\mb{M}^\mrT\mbV^{-1}\ba_i + \mb{M}^\mrT\mbV^{-1}\mb{M}) $ for some matrices $\mb{M}$ and $\mbV$. To complete the square, set $\mbV= \{ \bPsi_i^\mrT \mb{P}_i  \bPsi_i+ \mbQ \}^{-1}$ and match the coefficients of $\ba_i$, giving  $ \mbR^\mrT_i \mb{P}_i\bPsi_i = \mb{M}^\mrT\mbV^{-1} \implies \mb{M}  = \mbV \bPsi_i^\mrT \mb{P}_i \mbR_i$.
Thus, the full conditional posterior for $\ba_i$ is given by
$$\ba_i | \cdot \sim \mrN( \mb{\mu}_{\ba}, \mbV_{\ba} )$$ 
$$ \text{for } \mb{\mu}_{\ba} = \{ \bPsi_i^\mrT \mb{P}_i  \bPsi_i+ \mbQ \}^{-1} \bPsi_i^\mrT \mb{P}_i (\mbW_i-\mbU_i \bb) \text{ and }   \mbV_{\ba}= \{ \bPsi_i^\mrT \mb{P}_i  \bPsi_i+ \mbQ \}^{-1}.$$

\subsection*{Random effects precision matrix}
Assume the $\mrm \times 1$ vector $\ba_i|\mbQ \sim {\mrN}_\mrm(0, \mbQinv)$ for $i=1,\hdots,\mrN$ and the $\mrm \times \mrm$ precision matrix $\mbQ \sim \mr{Wishart}_{\mrm}(\bV, \nu)$ for which the kernel of the density is given by $ |\mbQ|^{(\nu - \mrm - 1)/2} \mr{exp} \left \{ -\frac{1}{2} \mr{tr}(\mb{V}^{-1}\mbQ)  \right \} $.
Define $\mb{S}= \sum_{i=1}^\mrN{\ba_i\ba_i^\mrT }$ as the sum of squares matrix of $\ba_i$. We use $\mb{S}$ to write 
$\sum_{i=1}^\mrN{\ba_i^\mrT \mbQ \ba_i}=\mr{tr}(\sum_{i=1}^\mrN{\ba_i^\mrT \mbQ \ba_i})= \mr{tr}(\sum_{i=1}^\mrN{\ba_i\ba_i^\mrT \mbQ }) = \mr{tr}(\mb{S}\mbQ)$ in the kernel of the multivariate normal density, using the properties $\mr{tr}(a)=a$ for scalar $a$ and $\mr{tr}(\mb{A}\mb{B})= \mr{tr}(\mb{B}\mb{A})$.
We also use the following properties of the trace to combine like-terms: 1) $|\bm{A}|^{-1}=|\bm{A}^{-1}|$ for $\bm{A}$ invertible and 2) for two square matrices $\bm{A}$ and $\bm{B}$ of the same dimension, $\mr{tr}(\bm{A}+\bm{B})=\mr{tr}(\bm{A})+\mr{tr}(\bm{B})$.
Using this, we can show the conditional posterior $p(\mbQ| \cdot)$ for $\mbQ$ is proportional to the kernel of a $\mr{Wishart}_{\mrm}\{ (\mb{S}+\bV^{-1})^{-1}, \mrN+ \nu\}$:
\begin{align}
\label{postQ}
p(\mbQ| \cdot) & \propto \prod_{i=1}^\mrN p(\ba_i| \mbQ)\times p(\mbQ) \nonumber\\
& \propto |\mbQinv|^{-\mrN/2} \mr{exp} \left \{ -\frac{1}{2} \mr{tr}(\mb{S}\mbQ)  \right \}  \times 
|\mbQ|^{(\nu - \mrm - 1)/2} \mr{exp} \left \{ -\frac{1}{2} \mr{tr}(\mb{V}^{-1}\mbQ)  \right \}  \\
&  \propto |\mbQ|^ {(\mrN+ \nu - \mrm - 1)/2}  \mr{exp} \left [ -\frac{1}{2} \mr{tr} \{ (\mb{S}+\mb{V}^{-1})\mbQinv \}  \right ]. \nonumber
\end{align}
Thus, $\mbQ| \cdot \sim \mr{Wishart}_{\mrm}\{ (\mb{S}+\bV^{-1})^{-1}, \mrN+ \nu\}$.

\subsection*{Fixed effects}

%Assume as before that the $2\mrL_i \times 1$ response vector for subject $i$ is $\mbW_i|\ba_i \overset{indep}{\sim} \mrN_{2 \mrL_i} \big(\mbU_i \bb + \bPsi_i \ba_i, \mb{D}_i\big)$ for $\mb{D}_i=\mr{diag}(\tau^2_\c,1) \otimes \mb{\mr{I}}_{\mrL_i} $, or in terms of the precision, $\mb{P}_i=\mb{D}_i^{-1}=\mr{diag}(\omega_\c, 1) \otimes \mb{\mr{I}}_{\mrL_i} $. Define the $2\mrL_i \times 1$ vector $\mbU_i=\mbW_i - \bPsi_i \ba_i$. The $\mrr \times 1$ vector $\bb \sim \mrN_\mrr(\mb{0}, \mbC^{-1})$ for $ \mbC^{-1}=\sigma^2_{\bb} \bI_\mrr$. To find the full conditional distribution of $\bb |\cdot$, we have 
\begin{align}
p(\bb | \cdot) & \propto \prod_{i=1}^\mrN  p(\mbW_i | \cdot ) \times p(\bb)   \\
& \propto  \exp \left(  -\frac{1}{2}  \left[ \sum_{i=1}^\mrN \left \{
(  \mbU_i \bb -  \mbU_i )^\mrT \mb{P}_i (  \mbU_i \bb -  \mbU_i )  \right\}+ \bb^\mrT \mbC \bb     \right] \right )  \\
& \propto  \exp \left[  -\frac{1}{2}   \sum_{i=1}^\mrN \left \{
(  \mbU_i \bb -  \mbU_i )^\mrT \mb{P}_i (  \mbU_i \bb -  \mbU_i )  \right\}  -\frac{1}{2}   \bb^\mrT \mbC \bb     \right]  \\
& \propto  \exp \left[  -\frac{1}{2}   \sum_{i=1}^\mrN \left \{
\bb^\mrT \mbU_i^\mrT \mb{P}_i  \mbU_i \bb 
-2\mbU_i^\mrT\mb{P}_i \mbU_i \bb
+ \mbU_i^\mrT \mb{P}_i \mbU_i
\right\} -\frac{1}{2}   \bb^\mrT \mbC \bb \right]  \\
\intertext{\centering and ignoring constant terms results in}  
& \propto  \exp \left[  -\frac{1}{2}  \left \{\bb^\mrT \mbC \bb +
\bb^\mrT\left(\sum_{i=1}^\mrN \mbU_i^\mrT \mb{P}_i  \mbU_i \right)\bb 
-2 \left( \sum_{i=1}^\mrN  \mbU_i^\mrT\mb{P}_i \mbU_i \right) \bb
\right\}  \right]  \\
& \propto  \exp \left[  -\frac{1}{2}  \left \{
\bb^\mrT\left(\mbC + \sum_{i=1}^\mrN \mbU_i^\mrT \mb{P}_i  \mbU_i  \right)\bb 
-2 \left( \sum_{i=1}^\mrN  \mbU_i^\mrT\mb{P}_i \mbU_i \right) \bb
\right\}  \right]  
\end{align}

As we did before, we want to form this term into the kernel of a Gaussian distribution and where the exponent is $ -1/2(\bb - \mb{M})^\mrT \mbV^{-1}  (\bb - \mb{M})= -1/2(\bb^\mrT \mbV^{-1} \bb - 2\mb{M}^\mrT\mbV^{-1}\bb + \mb{M}^\mrT\mbV^{-1}\mb{M}) $ for some matrices $\mb{M}$ and $\mbV$. To complete the square, set $\mbV= \left(\mbC + \sum_{i=1}^\mrN \mbU_i^\mrT \mb{P}_i  \mbU_i  \right)^{-1}$ and match the coefficients of $\bb$, giving  $\sum_{i=1}^\mrN  \mbU_i^\mrT\mb{P}_i \mbU_i = \mb{M}^\mrT\mbV^{-1} \implies \mb{M}  = \mbV \left( \sum_{i=1}^\mrN  \mbU_i^\mrT\mb{P}_i \mbU_i \right)^\mrT$.
Thus, the full conditional posterior for $\bb$ is given by
$$\bb | \cdot \sim \mrN( \mb{\mu}_{\bb}, \mbV_{\bb} )$$ 
$$ \text{for }  \mb{\mu}_{\bb}  =   \mbV_{\bb} \left\{ \sum_{i=1}^\mrN  (\mbW_i - \bPsi_i \ba_i )^\mrT\mb{P}_i \mbU_i \right \}^\mrT
\text{ and }  \mbV_{\bb}= \left( \sigma^{-2}_{\bb}\bI_\mrr + \sum_{i=1}^\mrN \mbU_i^\mrT \mb{P}_i  \mbU_i  \right)^{-1}.$$

\subsection*{Error Variance (Precision)}

Assume that the error precision $\omega_\c \sim \mr{Gamma}(g,h)$ where we parameterize the density such that $g$ is the shape parameter and $h=1/s$ is the inverse of the scale parameter, called the rate parameter. Specifically, if $X \sim \mr{Gamma}(g,h)$ then $ p(x|g,h) = x^{g-1} e^{-xh} \{h^g/ \Gamma(g)\}$. For simplicity of notation, denote the continuous response at $t_\ell$ for subject $i$  as $Y_{i\ell}=Y_{\c i}(t_\ell)$ for $i=1,\hdots,\mrN$ and $\ell= 1,\hdots,{\mrL_i}$, and define the total number of responses observed as $n=\sum_{i=1}^\mrN{\mrL_i}$. Let $Y_{i\ell} \overset{indep}{\sim} \mrN(\mu_{i\ell}, \mr{precision}= \omega_\c  )$. Then 
\begin{align}
p(\omega_\c | \cdot) & \propto \prod_{i=1}^\mrN \prod_{\ell=1}^{{\mrL_i}} p( Y_{i\ell} | \cdot ) \times p(\omega_\c)   \\
& \propto    \omega_\c^{n/2}     \exp \left\{ -\frac{\omega_\c}{2} \sum_{i=1}^\mrN \sum_{\ell=1}^{{\mrL_i}} ( Y_{i\ell} - \mu_{i\ell} )^2   \right \}  \times
\omega_\c^{g-1} \exp(- \omega_\c h)\\
& \propto    \omega_\c^{(n/2+g)-1}     \exp \left[ -\omega_\c \left \{
\frac{1}{2} \sum_{i=1}^\mrN \sum_{\ell=1}^{{\mrL_i}} ( Y_{i\ell} - \mu_{i\ell} )^2 + h  \right \} \right ]. 
\end{align}
This is the kernel of a Gamma density, so the posterior for $\omega_\c | \cdot \sim \mr{Gamma}(g_\omega, h_\omega)$ with shape and rate parameters	$g_\omega =n/2+g$ and $h_\omega= 1/2 \sum_{i=1}^\mrN \sum_{\ell=1}^{{\mrL_i}} ( Y_{i\ell} - \mu_{i\ell} )^2 + h $.

\end{document}